# A Note on "A survey of preference estimation with unobserved choice set heterogeneity" by Gregory S. Crawford, Rachel Griffith, and Alessandro Iaria.

C. Angelo Guevara, Universidad de Chile

## Abstract

*Crawford's et al. (2021) article on estimation of discrete choice models with unobserved or latent consideration sets, presents a unified framework to address the problem in practice by using "sufficient sets", defined as a combination of past observed choices. The proposed approach is sustained in a re-interpretation of a consistency result by McFadden (1978) for the problem of sampling of alternatives, but the usage of that result in Crawford et al. (2021) is imprecise in an important matter. It is stated that consistency would be attained if any subset of the true consideration set is used for estimation, but McFadden (1978) shows that, in general, one needs to do a sampling correction that depends on the protocol used to draw the choice set. This note derives the sampling correction that is required when the choice set for estimation is built from past choices. Then, it formalizes the conditions under which such correction would fulfill the uniform condition property and can therefore be ignored when building practical estimators, such as the ones analyzed by Crawford et al. (2021).*

Crawford et al. (2021) study the problem of choice sequences but, to simplify the notation and exposition of the issue, this note will refer to the modeling of a single choice in a traditional discrete choice model. The note begins defining the choice model, later revising McFadden's (1978) work on sampling of alternatives, and finally showing how it can be adapted to tackle the problem of unobserved or latent consideration sets and how it relates to Crawford's et al. (2021) results.

Consider first a Random Utility Model (RUM) setting in which the utility $U_{in}$ that an individual $n$ retrieves from alternative $i$ can be written as the sum of a systematic part $V_{in}$ and a random error term $\varepsilon_{in}$, as shown in Eq. (1)

$$U_{in} = V_{in} + \varepsilon_{in} = V(x_{in}, \beta^*) + \varepsilon_{in}, \qquad (1)$$

where $V_{in}$ depends on attributes $x_{in}$ and population parameters $\beta^*$.

Then, if $\varepsilon_{in}$ is distributed *iid* Extreme Value I $(0,\mu)$, the probability that $n$ will choose alternative $i$ will correspond to the Logit model shown in Eq. (2), where $C_n$ is the true consideration-set of $J_n$ elements from which individual $n$ selects one alternative. The scale $\mu$ in Eq. (2) is not identifiable and is thus usually normalized to equal 1.

$$P_n(i) = \frac{e^{\mu V_{in}}}{\sum_{j \in C_n} e^{\mu V_{jn}}} \qquad (2)$$

The problem originally tackled by McFadden (1978) was that the true consideration set $C_n$ was too large to be processed by the researcher in practice, what is solved by building a reduced practical set $D_n \subseteq C_n$ for estimation, using some sampling protocol conditional on the chosen alternative. Formally, $\pi(D_n | j)$ corresponds to the conditional probability that the researcher would sample a reduced set $D_n$, given that alternative $j$ was chosen by individual $n$. Under this setting McFadden (1978) showed that maximizing a pseudo-loglikelihood using the choice probabilities shown in Eq. (3) one can obtain consistent estimators of the population parameters $\beta^*$.

$$\pi(i | D_n) = \frac{e^{V_{in} + \ln \pi(D_n|i)}}{\sum_{j \in D_n} e^{V_{jn} + \ln \pi(D_n|j)}} \qquad (3)$$

The virtue of the pseudo-likelihood depicted in Eq. (3) is that it only depends on the alternatives of the reduced set $D_n$, transforming a problem of possibly millions of alternatives, to one that just has a few. Besides, the resulting model has a closed form that corresponds to a simple Logit model with a correction term by alternative $\ln \pi(D_n | j)$ that only depends on the sampling protocol. This result holds thanks to the IIA property of the Logit model, but was extended to more flexible models like MEV (GEV), Logit Mixture and RRM, by Guevara and Ben-Akiva (2013 a,b) and Guevara et al. (2016), respectively.

When referring to McFadden's (1978) result, Crawford et al. (2021, Eq. 2.4) show a version of Eq. (3) in which the sampling correction $\ln \pi(D_n | j)$ is not included and it is stated that such curtailed model will achieve consistency as long as $D_n \subseteq C_n$. That statement is, in general, incorrect. It would only be valid for some special cases in which what McFadden called the "uniform conditioning property" holds, when $\ln \pi(D_n | j) = \ln \pi(D_n | j')$ so that the correction cancels out in Eq. (3). The uniform condition property holds, for example, if the full choice set has $J$ alternatives, and the sampling protocol corresponds to draw the chosen alternative and then to randomly draw $\tilde{J}-1$ nonchosen alternatives among the $J-1$ available, with the same probability *p*. In that case $\pi(D_n | j) = \binom{J-1}{\tilde{J}-1}^{-1}$ for all $j \in C_n$ and therefore cancels out in Eq. (3). As it will be shown, the uniform conditioning property does not necessarily hold for the problem studied by Crawford et al. (2021), but it may do it under some conditions.

Consider now a different challenge, the problem that the true consideration-set $C_n$ is latent to the researcher, but that she can observe the individual making *R* choices from $C_n$ in past instances. The researcher is interested in modeling the choice occurring at the instance *R+1*. For this, she builds a practical consideration set (or sufficient set in Crawford's 2021 words) that includes all the alternatives that were observed in the previous *R* instances, plus the alternative chosen on instance *R+1*, if it was not already included. This coincides with Crawford's et al. (2021) Past Purchase History (**PPH**) sufficient set, with the only differences that *T* in Crawford et al. (2020) corresponds to *R+1* here, and that in this case only the last choice is modeled. Other sufficient sets of this kind studied by Crawford et al. (2021), like the Full Purchase History (FPH) and the Choice Permutation (CP), collapse to PPH when only the last choice is considered. In practice, data needed to build a PPH sufficient set can be gathered, for example, from a series of supermarket purchases or other types of passive data that can be available, for example, in transportation networks. These observations could correspond to choices of the same individual *n* or, assuming some type of group homogeneity, choices made by diverse individuals that faced the same choice situation, on what is called Inter-Personal (**IP**) sufficient set by Crawford et al. (2021), or the Cohort consideration set by Arriagada et al. (2021).

The link between McFadden´s (1978) problem of sampling of alternatives and the latent consideration set issue requires making some **invariability assumptions**. First, if one can assume that the **choice behavior** and the **consideration set** do not vary across the *R+1* choices, past choices could be understood as draws with replacement from the true but latent consideration set. By this, the data generation process for the latent consideration set problem will precisely mimic the problem of estimation and sampling of alternatives studied by McFadden (1978). The protocol in this case will correspond to importance sampling with replacement, in which the sampling probability of each alternative corresponds to the choice probability $P_n(i|C_n, r)$ on instance $r$.

The behavioral and consideration set invariability assumption seems easy to sustain in the case of the PPH, if the past choices are "sufficiently recent" to not be prone to a contextual or behavioral changes. In turn, these assumptions may be harder to sustain in the case of IP but does not seem implausible.

The problem that remains is to determine the sampling correction needed under this setting, which is not trivial, not only because of the form that it takes, but also because it depends on the choice probabilities $P_n(i|C_n, r)$, which depend on the unknowns $C_n$ and $\beta^*$. To solve this dilemma, we need to make an additional assumption.

The sampling correction can get very complicated depending on the protocol considered. McFadden (1978) explores four cases to show some practical examples in which neglecting this correction results in inconsistent estimators. This effort is later deepened by Ben-Akiva and Lerman (1985; Ch.9) and then refined by Ben-Akiva (1989). Among other cases, the latter work derived the correction needed for the following sampling protocol: Draw with replacement $R$ times from the set of all alternatives $C$ with selection probabilities $q_j$, with $\sum_{j \in C} q_j = 1$, and add the chosen alternative if it was not already drawn. This protocol is closely related to the latent consideration set problem studied in this note, but to be able to apply it, we need to add a third type of **invariability** assumption such that $P_n(i|r) = P_n(i) = q_j$, which would be achieved if the **attributes** do not vary over the past choices. In such a case, the sampling correction needed will correspond to Eq. (4), where $n_j$ is the number of times alternative $j$ was chosen in the $R+1$ instances.

$$\pi(D_n | i) = \frac{n_i}{P_n(i)} \left( \frac{R!}{\prod_{j \in D} n_j!} \prod_{j \in D} P_n(j)^{n_j} \right) = \frac{n_i}{P_n(i)} K_D, \qquad (4)$$

The attributes' invariability assumption likely holds for models bases on supermarket data for sufficiently recent periods in which the attributes shall not change significantly. IN turn, the validity of this assumption may be more debatable in models based on transportation data, which are prone to day-to-day variability in, e.g., travel times.

Note now that the term $K_D = \frac{R!}{\prod_{j \in D} n_j!} \prod_{j \in D} P_n(j)^{n_j}$ on the right of Eq. (4), although complicated and depending on $P_n(j)$, it does not depend on the alternative $i$, and can therefore be omitted from the analysis since it would cancel out in Eq. (3). Besides, as the number of past choices $R$ grows, the sample error of the empirical probability will vanish, $n_i/R$ will become closer to the choice probability $P_n(i)$, resulting in that the whole sampling correction in Eq. (4) does not depend on the alternative $i$ and, therefore, will cancel out in Eq. (3) and can be ignored.

$$\pi(D_n | i) = K_D \frac{n_i}{P_i(i)} \approx K_D \frac{n_i}{n_i/R} = K_D R \tag{5}$$

Thus, the "sufficient set" approach to the problem of latent consideration sets will achieve consistency if the behavioral, consideration, and attributes' invariability assumptions hold, and the sample error of the empirical probability is small enough. As it is often the case, this implies a tradeoff. For example, on the one hand, it would be recommendable to build the sufficient set with recent enough choices to fulfill the behavioral, consideration and attributes set invariability, but also with old enough choices to ensure a good approximation of the choice probability. How and when a sufficient set will properly fulfil these conditions should be analyzed on a case-by-case basis.